\PassOptionsToPackage{compatibility=false}{caption}

\documentclass[10 pt]{article}

\usepackage[utf8]{inputenc}
\usepackage[american]{babel}
\usepackage{csquotes}
\usepackage[letterpaper, total={6in, 8in}]{geometry}
\usepackage{comment}
\usepackage{url}
\usepackage{hyperref}
\usepackage{graphicx}
\usepackage{subcaption}
\usepackage{multicol}
\usepackage{xspace}
\usepackage{amsmath, amssymb, amsfonts}
\usepackage{ntheorem}
\usepackage{mathtools}
\usepackage{multirow}
\usepackage{tabularx}

\newcolumntype{C}{>{\centering\arraybackslash}X}
\newcolumntype{L}{>{\raggedright\arraybackslash}X}
\usepackage{siunitx}
\usepackage{listings}
\usepackage{float}

\usepackage{pgfplots}
\usepgfplotslibrary{fillbetween}
\pgfplotsset{compat = newest}
\usetikzlibrary{arrows,positioning,shapes,intersections,patterns,calc,fit,external,decorations,decorations.markings} 
\newtheorem{definition}{Definition}
\newtheorem{lemma}{Lemma}
\newtheorem{theorem}{Theorem}

\newtheorem{remarkth}[definition]{Remark}
\newenvironment{remark}{\begin{remarkth}\upshape}{\end{remarkth}}

\newcommand{\N}{\mathbb{N}}

\usepackage{tikz}
\usetikzlibrary{matrix,arrows,calc,positioning,shapes,decorations.pathreplacing}
\tikzstyle{vertex}=[circle,fill=black!20,minimum size=15pt,inner sep=0pt]
\tikzstyle{selected vertex} = [vertex, fill=red!24]
\tikzstyle{edge} = [draw,thick,-]
\tikzstyle{dedge} = [draw,thick,<->]
\tikzstyle{shadowdedge} = [draw, dotted,->]
\tikzstyle{weight} = [font=\small]
\tikzstyle{selected edge} = [draw,line width=5pt,-,red!50]
\tikzstyle{ignored edge} = [draw,line width=5pt,-,black!20]
\pgfplotsset{select coords between index/.style 2 args={
    x filter/.code={
        \ifnum\coordindex<#1\fi
        \ifnum\coordindex>#2\fi
    }
}}
\usepackage{booktabs}

\newtheorem{assum}{Assumption}

\newcommand\tran{\mkern-2mu\raise1.25ex\hbox{$\scriptscriptstyle\top\hspace{0.5mm}$}\mkern-3.5mu}
\newcommand{\R}{\mathbb{R}}

\newcommand{\D}{\mathcal{D}}

\newcommand{\bm}[1]{{\boldsymbol{#1}}}

\DeclareMathOperator{\diag}{diag}
\DeclareMathOperator{\var}{var}

\DeclareMathOperator{\rank}{rank}

\DeclareMathOperator{\Prob}{P}
\newcommand{\GP}{\mathcal{GP}}

\newcommand{\x}{\bm x}

\newcommand{\q}{\bm q}

\newcommand{\f}{\bm{f}}

\renewcommand{\u}{\bm{u}}
\newcommand{\y}{\bm{y}}

\usepackage[noabbrev]{cleveref} 
\crefname{rem}{Remark}{Remarks}
\crefname{exam}{Example}{Examples}
\crefname{assum}{Assumption}{Assumptions}
\crefname{prop}{Proposition}{Propositions}
\crefname{propy}{Property}{Properties}
\crefname{cor}{Corollary}{Corollaries}
\crefname{lem}{Lemma}{Lemmas}
\crefname{section}{Section}{Sections}
\crefname{subsection}{Section}{Sections}
\crefname{thm}{Theorem}{Theorems}
\crefname{defn}{Definition}{Definitions}
\crefname{figure}{Fig.}{Fig.}
\Crefname{figure}{Figure}{Figures}
\crefname{equation}{}{}

\begin{document}

\title{\LARGE \bf Safe Online Learning-based Formation Control of\\ Multi-Agent Systems with Gaussian Processes}
\author{Thomas Beckers$^{1}$, Sandra Hirche$^{2}$, and Leonardo Colombo$^{3}$%
\thanks{Preprint submitted to IEEE CDC 2021}%
}
\maketitle
\thispagestyle{empty}
\pagestyle{empty}

\author{$^{1}$ Department of Electrical and Systems Engineering, University of Pennsylvania,\\\phantom{aaaa} Philadelphia, USA. {\tt\small tbeckers@seas.upenn.edu}

\and $^{2}$ Department of Electrical and Computer Engineering, Technical University of Munich,\\\phantom{aaaa}  Munich, Germany. {\tt\small hirche@lsr.ei.tum.de}

\and $^{3}$ Instituto de Ciencias Matemáticas (CSIC-UAM-UCM-UC3M), Madrid, Spain.\\\phantom{aaaa}  {\tt\small leo.colombo@icmat.es}
}

%

\maketitle
 \thispagestyle{empty}
 \pagestyle{empty}

\begin{abstract}
Formation control algorithms for multi-agent systems have gained much attention in the recent years due to the increasing amount of mobile and aerial robotic swarms. The design of safe controllers for these vehicles is a substantial aspect for an increasing range of application domains. However, parts of the vehicle's dynamics and external disturbances are often unknown or very time-consuming to model. To overcome this issue, we present a safe formation control law for multiagent systems based on double integrator dynamics by using Gaussian Processes for an online learning of the unknown dynamics. The presented approach guarantees a bounded error to desired formations with high probability, where the bound is explicitly given. A numerical example highlights the effectiveness of the learning-based formation control law.
\end{abstract}

\section{Introduction}\label{sec1}
Extending the concept of a single autonomous mobile robot performing a
   task to a group of robots has been an area
  of active research in last decades.
  One of the key elements in the operation of groups of mobile robots that require a specific spatial configuration is the control method used to coordinate the behavior of each robot~\cite{jorgebook}.\\
Some of the most widely used concepts are based on virtual potential fields that generate attraction and repulsion forces between robots~\cite{olfati2002distributed}, or to react to objects in the environment~\cite{balch:behaviorbased}. Another proposed method makes use of the concept of a virtual articulated mechanism that can be rotated, deformed and scaled over time~\cite{kittsmas:mecha}. Others, based on the concept of swarms of self organized robot groups~\cite{belta:swarm_abstraction,freeman:swarms}, rely on the idea
  of reducing the problem to a lower dimensional representation to allow for
  scaling and abstraction. This fact provides a significant benefit for the control of groups with a
  great number of agents, but it is not well-suited for missions that require
  to take into account uncertainties or external disturbances in the dynamics of the agents to reach a desired collective behaviour.\\
  Formation control strategies are a powerful tool for the coordination of a team of robots in multi-robot systems  where spatial constraints are defined among agents, as surveyed by~\cite{oh2015survey}.  However, these control approaches depend on exact models of the systems and exclude the effect of external disturbances in order to guarantee stability of the agents to a desired formation shape. In this paper we will show how to tackle the formation control problem when a multi-agent system is formulated under a double integrator dynamics and subject to unknown dynamics, in the case where agents only measure inter-agent distances.\\
  One of the most popular distance-based formation stabilization techniques are the ones based on the negative gradient of potential functions~\cite{dika} and also by rigidity theory~\cite{asimov}. In the case of rigidity theory, most of techniques to stabilize a set of agents to a desired shape using distance measurements are given for single integrator agents~\cite{helmke2013equivariant},~\cite{krick2009stabilisation}. Double integrator models have been extensively studied for flocking control since the pioneering works~\cite{olfati2006flocking} and~\cite{tanner2003stable}, but not for distance-based formation control, since a full characterization of the convergence analysis is still a challenging problem in the  multi-agent systems literature~\cite{sun2015rigid}. To overcome the issue of unknown dynamics, learning-based control laws have been proposed but they are limited to iterative learning~\cite{liu2012iterative}, leader-follower formations~\cite{yuan2017formation,8870252} or lack of guarantees~\cite{8172432}. To the best of the authors' knowledge, there are no available results for the design of a safe distance-based formation control law for double integrator agents under partially unknown dynamics, based on online learning data-driven models.\\
  These data-driven modeling tools have shown remarkable results in many different applications including control, machine learning and system identification~\cite{brunton2019data}. In data-driven control, data of the unknown system dynamics is collected and used to predict the dynamics in areas without training data. In contrast to parametric models, those models are highly flexible and are able to reproduce a large class of different dynamics, see~\cite{hou2013model}.\\
  Within the past two decades, Gaussian process (GP) models~\cite{rasmussen2006gaussian} has been increasingly used for modeling dynamical system due to many beneficial properties such as the bias-variance trade-off and the strong connection to Bayesian mathematics. In contrast to many other techniques, GP models provide not only a prediction but also a measure for the uncertainty of the model. This powerful property makes them very attractive for many applications in control, e.g., model predictive control~\cite{hewing2019cautious}, feedback linearization~\cite{umlauft:TAC2020}, and tracking control~\cite{beckers2019stable}, as the uncertainty measure allows to provide performance and safety guarantees. Recently, GP models have been employed for the control of multi-agent systems in~\cite{le2020gaussian,yang2021distributed} but without convergence guarantees or online learning.\\
  The purpose of this article is to employ the power of learning-based approaches, in particular, Gaussian processes, for the formation control of multi-agent systems with partially unknown double integrator dynamics. The main contribution of this article is a safe online learning-based formation control law for this class of multi-agent systems. The proposed decentralized control law guarantees the boundedness of the error to desired formation shapes, and specifies the ultimate bound.\\
  We begin this work by reviewing the necessary background and definitions about GPs and rigidity theory for multi-agent formations in Section \ref{sec2}. Section \ref{sec3} addresses the problem of modeling the partially unknown dynamics with GPs and the model error. Section \ref{sec4} describes the online learning-based control law and the bounded error to desired formation shapes. Finally, a numerical example is presented in Section \ref{sec5}.\\
  \textbf{Notation:} Vectors are denoted with bold characters and matrices with capital letters. The expression~$\mathcal{N}(\mu,\Sigma)$ describes a normal distribution with mean~$\mu$ and covariance~$\Sigma$. The probability function is denoted by $\Prob$. $\mathbb{E}[X]$ denotes the expected value of a random variable $X$. The set $\R_{>0}$ denotes the set of positive real numbers. The Euclidean norm is denoted by $||\cdot||$ and by $|\mathcal{X}|$ the cardinal of the set $\mathcal{X}$.
\section{Background and Definitions}\label{sec2}
\label{sec: LA}
We begin by introducing the necessary background on GP models (see~\cite{rasmussen2006gaussian} for more details) and the basics of rigidity theory for formation control (see~\cite{asimov} for more details).
\subsection{Gaussian Process models}
\label{sectionGPR}
A Gaussian process is a stochastic process which is completely defined by a mean function $m_{\mathrm{GP}}: \R^p \rightarrow \R$ and a kernel function $k: \R^p \times \R^p \rightarrow \R$. One of the main advantages of GPs is their combined use with Bayes' Theorem to provide statistical inference with function regression~\cite{rasmussen2006gaussian}. In this regard, consider the output $\bm{y}$ of a function $\bm{f}: \R^{p} \rightarrow \R^{p}$, where
\begin{align}
\bm f(\x)&=
\begin{cases} 
f_1(\x)\sim \GP(\bm{0},k(\x,\x^\prime))\\
\vdots\hspace{0.9cm}\vdots\hspace{0.5cm}\vdots\\
f_p(\x)\sim \GP(\bm{0},k(\x,\x^\prime)).
\end{cases}
\end{align}
The mean function is set to zero without loss of generality. The measurements might be affected by Gaussian noise such that $\bm{y} = \bm{f}(\x) + \bm{\eta}$, where $\bm{\eta}\sim\mathcal{N}(0,\sigma^2 I_p)$ with the $p$-dimensional identity matrix $I_p$. The training set $ \mathcal{D=} \left\lbrace X,Y \right \rbrace$ denotes the set of input data, $X=\left[\x^{\{1\}}, \x^{\{2\}}, \ldots, \x^{\{m\}}\right] \in \R^{p \times m}$ and measured output data, $Y=\left[\bm{y}^{\{1\}}, \bm{y}^{\{2\}}, \ldots, \bm{y}^{\{m\}}\right] \in \R^{p \times m}$. For a test input $\x^{*}\in\R^p$, the predictions of $\bm{f}(\x^{*})$ are provided by conditioning on the data which leads to the posterior distribution 
\begin{align}
\mu\left(f_{i}\!\mid\!\x^{*}, \D\right)&\!=\! \bm{k}\left(\x^{*}, X\right)^{\!\top}\!\left(K+I \sigma^{2}\right)^{-1} Y_{:, i},\\
    \var\left(f_{i}\mid\x^{*}, \D\right)&= k\left(\x^{*}, \x^{*}\right)-\bm{k}\left(\x^{*}, X\right)^{\top}\notag\\
    &\phantom{=}\left(K+I\sigma^{2}\right)^{-1} \bm{k}\left(\x^{*}, X\right)\notag
\end{align}
for all $i\in\{1,\ldots,p\}$, where $Y_{:,i}$ denotes the $i$-th column of the matrix of outputs~$Y$. The kernel $k$ is a measure for the correlation of two inputs~$(\x,\x^\prime)$. The function~$K\colon \R^{p\times m}\times  \R^{p\times m}\to\R^{m\times m}$ is called the Gram matrix whose elements are given by $K_{j',j}= k(X_{:, j'},X_{:, j})+\delta(j,j')\sigma^2$ for all $j',j\in\{1,\ldots,m\}$ with the delta function $\delta(j,j')=1$ for $j=j'$ and zero, otherwise. The vector-valued function~$\bm{k}\colon \R^p\times  \R^{p\times m}\to\R^m$, with the elements~$k_j = k(\x^*,X_{:, j})$ for all $j\in\{1,\ldots,m\}$, expresses the covariance between~$\x^*$ and the input training data $X$. The selection of the kernel and the determination of the corresponding hyperparameters can be seen as degrees of freedom of the regression. A powerful kernel for GP models of physical systems is the squared exponential kernel. An overview about the properties of different kernels can be found in~\cite{rasmussen2006gaussian}. 
\begin{remark}
For simplicity, we consider identical kernels for each output dimension. However, the GP model can be easily adapted to different kernels for each output dimension.
\end{remark}
\subsection{Rigidity of formations}\label{sec2b}
Consider $n\geq 2$ autonomous agents whose positions are denoted by $\bm{p}_i\in\R^{d}$, $d=\{2,3\}$ and denote by $\bm{p}\in\R^{dn}$ the stacked vector of agents' positions.\\
The neighbor relationships between agents are described by an undirected graph $\mathbb{G} = (\mathcal{N}, \mathcal{E})$ with the ordered edge set $\mathcal{E}\subseteq\mathcal{N}\times\mathcal{N}$. The set of neighbors for $i\in\mathcal{N}$, denoted by $\mathcal{N}_i$, is defined by $\mathcal{N}_i:=\{j\in\mathcal{N}:(i,j)\in\mathcal{E}\}$. Agents can sense the relative positions of its nearest neighbors, in particular, agents can measure its Euclidean distance from other agents in the subset $\mathcal{N}_i \subseteq \mathcal{N}$. We define the elements of the incidence matrix $B\in\R^{|\mathcal{N}|\times|\mathcal{E}|}$ that establish the neighbors' relationships for $\mathbb{G}$ by
\begin{align}
	B_{i,k} = \begin{cases}+1  &\text{if} \quad i = {\mathcal{E}_k^{\text{tail}}} \\
		-1  &\text{if} \quad i = {\mathcal{E}_k^{\text{head}}} \\
		0  &\text{otherwise}
	\end{cases},
	\label{eq: B}
\end{align}
where $\mathcal{E}_k^{\text{tail}}$ and $\mathcal{E}_k^{\text{head}}$ denote the tail and head nodes, respectively, of the edge $\mathcal{E}_k$, i.e., $\mathcal{E}_k = (\mathcal{E}_k^{\text{tail}},\mathcal{E}_k^{\text{head}})$. The stacked vector of relative positions between neighboring agents, denoted by $\bm{z}\in\R^{d|\mathcal{N}|}$, is given by
\begin{align}
	\bm{z} = \overline B^T \bm{p},
\end{align}
where $\overline B := B \otimes I_{d}\in\R^{d|\mathcal{N}|\times d|\mathcal{E}|}$ with $I_{d}$ being the $(d\times d)$ identity matrix, and $\otimes$ the Kronecker product. Note that $\bm{z}_k \in \R^d$ and $\bm{z}_{k+|\mathcal{E}|}\in\R^d$ in $\bm{z}$ correspond to $\bm{p}_i - \bm{p}_j$ and $\bm{p}_j - \bm{p}_i$ for the edge $\mathcal{E}_k$. We can also define $\bm{p}_{ij} := \bm{p}_i - \bm{p}_j$ to reduce unnecessary verbosity. A \textit{framework} for $\mathbb{G}$ is then defined as the pair $(\mathbb{G},\bm{p})$.\\
In this work, conditions to guarantee convergence to desired formations are based on the property called \textit{rigidity} of the desired formation shape. According to this, the \textit{rigidity matrix} for the framework $(\mathbb{G},\bm{p})$ is defined as (see \cite{asimov} for instance)
\begin{align}
    R(\bm{z})=\frac{1}{2} \frac{\partial \ell_{\mathbb{G}}(\bm{p})}  {\partial \bm{p}} = D(\bm{z})^{\top} \overline{B} \in \R^{|\mathcal{E}| \times d|\mathcal{N}|},
\end{align}
 with $D(\bm{z}) = \diag(\bm{z}_{1}, \ldots,\bm{z}_{|\mathcal{E}|}) \in \R^{d|\mathcal{E}| \times |\mathcal{E}|}$ and distance measure function $\ell_{\mathbb{G}}: \R^{d|\mathcal{N}|} \rightarrow \R^{|\mathcal{E}|}$ defined by 
\begin{align}
     \ell_{\mathbb{G}}(\bm{p})=\left(\left\|\bm{p}_{i}-\bm{p}_{j}\right\|^{2}\right)_{(i, j) \in \mathcal{E}}=D^\top(\bm{z})\bm{z}.
\end{align}
We consider the desired distance between neighboring agents over the edge $\mathcal{E}_k$ as $d_{k}$ and we further define the squared distance error for the edge $\mathcal{E}_k$ as
\begin{align}\label{e_k}
    \bm{e}_{k}=\left\|\bm{p}_{i}-\bm{p}_{j}\right\|^{2}-d_{k}^{2}=\left\|\bm{z}_{k}\right\|^{2}-d_{k}^{2},
\end{align} 
with the stacked squared distance vector error denoted by $\bm{e} = \left[\bm{e}_{1}, \ldots, \bm{e}_{|\mathcal{E}|}\right]^{\top}$. For $k\in\{1,\ldots,|\mathcal{E}|\}$, the set of desired shapes is then defined by
\begin{align}\label{S}
    \mathcal{S}=\{\bm{z}\in\R^{d|\mathcal{N|}}|\,\,||\bm{z}_k||=d_k \}.
\end{align}
A framework $(\mathbb{G},\bm{p})$ is said to be \textit{rigid} if it is not possible to smoothly move one node of the framework without moving the rest while maintaining the inter-agent distance given by $\ell_{\mathbb{G}}(\bm{p})$, see~\cite{asimov}. An \textit{infinitesimally rigid} framework is a rigid framework which is invariant under and only under infinitesimally transformations under $R(\bm{z})$, i.e., $\ell_{\mathbb{G}}(\bm{p}+\delta \bm{p})=\ell_{\mathbb{G}}(\bm{p})$ where $\delta \bm{p}$ denotes an infinitesimal displacement of $\bm{p}$.\\
It is well known (see \cite{asimov}) that a framework  $(\mathbb{G},\bm{p})$ is infinitesimally rigid in $\R^d$ if $\bm{p}$ is a regular value of  $\ell_{\mathbb{G}}(\bm{p})$ and $(\mathbb{G},\bm{p})$ is rigid in $\R^d$. In particular, $(\mathbb{G},\bm{p})$ is \textit{infinitesimally rigid} in $\R^2$ if $\rank R(\bm{z})=2n-3$ (respectively,  $\rank R(\bm{z})=3n-6$ in $\R^3$). The framework $(\mathbb{G},\bm{p})$ is said to be \textit{minimally rigid} if it has exactly $2n-3$ edges in $\R^2$ or $3n-6$ edges in $\R^{3}$. This means that if we remove one edge from a minimally rigid framework $(\mathbb{G},\bm{p})$, then it is not rigid anymore. Thus, the only motions over the agents in a minimally rigid framework, while they are already in the desired shape, are the ones defining translations and rotations of the whole shape, see \cite{oh2015survey}.\\
One important property for the stabilization to desired shapes in distance-based formation control is that the rigidity matrix $R(\bm{z})$ has full row rank if the framework $(\mathbb{G},\bm{p})$ is minimally and infinitesimally rigid, see \cite{asimov} for more details. 
\section{Modeling formation control of double integrator agents with Gaussian Processes}\label{sec3}
Consider the set $\mathcal{N}$ consisting of $n\geq 2$ free autonomous agents evolving on $\R^d$ with $d=\{2,3\}$ as in Section \ref{sec2b}, under a double integrator dynamics, that is
\begin{align}\label{double_wo}
	\begin{cases}
	\dot{\bm{p}} =\bm{v} \\
	\dot{\bm{v}} =\bm{u}.
	\end{cases}
\end{align}
By considering the control law $\u(t)=-\mathcal{K}\bm{v}-R^\top(\bm{z})e(\bm{z})$, the closed loop system is given by 
\begin{align}\label{formation}
	\begin{cases}
	\dot{\bm{p}} =\bm{v} \\
	\dot{\bm{v}} = -\mathcal{K}\bm{v}-R^\top(\bm{z})e(\bm{z}).
	\end{cases}
\end{align} with $\mathcal{K}=K\otimes I_{d}$ and $K$ the gain diagonal matrix with the $i$-th entry being $k_i>0$. The closed-loop system \cref{formation} is called double integrator formation stabilization system, see \cite{sun2015rigid} for instance. Note that the role of equations \cref{formation} is to stabilize a desired rigid shape and reach a stationary formation with zero velocities of the agents.\\
To reach a desired shape, for each edge $\mathcal{E}_k=(i,j)$ in the infinitesimally and minimally rigid framework we introduce the artificial potential functions $V_{k}:\R^{d}\to\R$, given by
\begin{align}
	V_{k}(\bm{z}_k)=\frac{1}{4}(||\bm{z}_{k}||^2-d_{k}^2)^2,
	\label{eq: Vij}
\end{align} to provide a measure for the interaction between agents and their nearest neighbors (see \cite{oh2015survey} for a detailed discussion on the choices of elastic potential functions).  In these potentials,~$\bm{z}_k$ denotes the relative position between agents for the edge $\mathcal{E}_k$, and $d_{k}$ denotes the desired length for the edge $\mathcal{E}_k$. Note also that the artificial potential (\ref{eq: Vij}) is not unique, and it can be given by other similar expressions as it was discussed by~\cite{oh2015survey}. Therefore, we can define the artificial potential function $V_0:\R^{d|\mathcal{N}|}\to\R$ for the overall networked control system as 
\begin{align}
    V_0(\bm{z})=\sum_{k=1}^{|\mathcal{E}|}V_k(\bm{z}_k).
\end{align}
In order to control the velocity of the agents, we consider the potential function $V_1:\R^{d|\mathcal{N}|}\to\R$ defined as
\begin{align}
    V_1(\bm{v})=\frac{1}{2}\sum_{i=1}^{|\mathcal{N}|}||\bm{v}_i||^2.
\end{align}
By considering $V_0+V_1$ as energy function of the networked control system with double integrator dynamics, one can show the local asymptotic convergence of the formation to the shape given by \cref{S} with velocity zero for all the agents if the framework $(\mathbb{G},\bm{p})$ is rigid \cite{ahn14} (respectively, local exponential stability for infinitesimally and minimally rigid frameworks, see \cite{sun2015rigid}).\\
Now, consider each agent $i\in\{1,\ldots,|\mathcal{N}|\}$ as double integrator system disturbed by an additive unknown dynamics given by \begin{align}\label{double}
    \begin{cases}
      \dot{\bm{p}}_i=\bm{v}_i, \\
      \dot{\bm{v}}_i= \u_i+\f_i(\bm{p}_i,\bm{v}_i),
    \end{cases}       
\end{align}
where $\f_i:\R^{2d}\to\R^{d}$ is a state-dependent unknown function. Note that the time-dependency of the states is omitted for simplicity of notation and the time dependency of the unknown input forces $\f_i$ might be also indirect, i.e. $\f_i(\bm{p}_i(t),\bm{v}_i(t))$. The goal is to find a decentralized control law $\bm{u}(t)\in\R^{d|\mathcal{N}|}$ to converge to a desired formation of an infinitesimally and minimally rigid framework.\\
In the following, we propose an online learning strategy and an upper bound for the error estimation between the learned (i.e., the mean prediction of the GP) and the true dynamics. In preparation for the learning and control step, we introduce the estimate $\hat{\f}_i\colon\R^{2d}\to\R^{d}$ which can include existing prior knowledge about the unknown dynamics $\f_i$, e.g., using  off-the-shelf modeling or classical system identification. However, if no prior knowledge is available, the estimate $\hat{\f}_i$ is set to zero. Then, the system of agents~\cref{double} can be rewritten as
 \begin{align}\label{double2}
    \begin{cases}
      \dot{\bm{p}}=\bm{v}, \\
      \dot{\bm{v}}=\u+\bm{\rho}(\bm{q})+\hat{\f}(\bm{p},\bm{v}),
    \end{cases}   
\end{align}
with the stacked vector of estimating functions $\hat{\f}(\bm{p},\bm{v})=[\hat{\f}_1(\bm{p}_1,\bm{v}_1)^\top,\ldots,\hat{\f}_{|\mathcal{N}|}(\bm{p}_{|\mathcal{N}|},\bm{v}_{|\mathcal{N}|})^\top]^\top$ and the unknown dynamics $\bm{\rho}\colon\R^{2d|\mathcal{N}|}\to\R^{d|\mathcal{N}|}$ with elements defined by
\begin{align}\label{estimation-error}
    \bm{\rho}_i(\q_i)=\f_i(\bm{p}_i,\bm{v}_i)-\hat{\f}_i(\bm{p}_i,\bm{v}_i).
\end{align}
where $\q_i=[\bm{p}_i^\top,\bm{v}_i^\top]^\top$. In the next step, we employ a GP model for the learning of the unknown dynamics $\bm{\rho}$. For this purpose, each agent collects $m(t)\in\N$ training points of based on its own dynamics~\cref{double} such that data sets
\begin{align}
    \D_{i,m(t)}=\{\bm{q}_i^{\{j\}},\y_i^{\{j\}}\}_{j=1}^{m(t)}\label{for:dataset}
\end{align}
are created. The output data $\y_i\in\R^d$ are given by $\y_i=\dot{\bm{v}}_i-\hat{\f}_i(\bm{p}_i,\bm{v}_i)-\u_i$. The number of training points $m(t)$ of the data sets $\D_{i,m(t)},i\in\{1,\ldots,|\mathcal{N}|\}$ with $m\colon \R_{\geq 0}\to\N$ can change over time $t$, i.e., it allows online learning. Let $\D_{m(t)}=\{\D_{i,m(t)},\ldots,\D_{|\mathcal{N}|,m(t)}\}$ be a set that contains all training set. Then, we introduce the following assumption on the data collection.
\begin{assum}\label{ass:1}
    There are only finitely many switches of~$m(t)$ over time and there exists a time $T\in\R_{\geq 0}$ where $\D_{m(T)}=\D_{m(t)},\forall t\geq T,\forall i\in\{1,\ldots,|\mathcal{N}|\}$.
\end{assum}
Assumption \ref{ass:1} is little restrictive since the number of sets is often naturally bounded due to finite computational power or memory limitations and since the unknown functions $\bm{\rho}$ in~\cref{double2} is not explicitly time-dependent, long-life learning is typically not required. Furthermore,~\cref{ass:1} ensures that the switching between the data sets is not infinitely fast which is natural in real world applications. To model the error, an assumption has to be made about the kernel function $k$ of the GP model. 
\begin{assum}\label{ass:2}
    Let the continuous kernel $k$ be chosen in such a way that the functions $\rho_i$, $i\in\{1,\ldots,d|\mathcal{N}|\}$ have a bounded reproducing kernel Hilbert Space (RKHS) norm on a compact set $\Omega \subset \R^{2d|\mathcal{N}|}, \text { i.e. }\left\|\rho_i\right\|_{k}<\infty \text { for all } i\in\{1, \ldots, d|\mathcal{N}|\}$. 
\end{assum}
The norm of a function in a RKHS is a smoothness measure relative to a kernel~$k$ that is uniquely connected with this RKHS. In particular, it is a Lipschitz constant with respect to the metric of the used kernel. A more detailed discussion about RKHS norms is given in~\cite{wahba1990spline}. Note that the previous assumption also requires that the kernel must be selected in such a way that the function $\bm{\rho}$ is an element of the associated RKHS. This sounds paradoxical since this function is unknown. However, there exist some kernels, namely universal kernels, which can approximate any continuous function arbitrarily precisely on a compact set~\cite[Lemma 4.55]{steinwart2008support} such that the bounded RKHS norm is a mild assumption.\\
Under the previous consideration on the model under study, the model error can be bounded as written in the following lemma.
\begin{lemma}\label{model-error}Consider the system \cref{double2} and a GP model satisfying~\cref{ass:1,ass:2}. Then the model error is probabilistically bounded by 
$$\Prob\left\{\|\bm{\mu}(\bm{\rho} \mid \bm{q}, \D_m)-\bm{\rho}(\bm{q})\| \leq\left\|\bm{\beta}^{\top} \Sigma^{\frac{1}{2}}(\bm{\rho} \mid \bm{q}, \D_m)\right\|\right\} \geq \delta$$ for $\bm{q} \in \Omega\subset\R^{2d|\mathcal{N}|}$ compact, with  $\delta \in(0,1), \bm{\beta}, \bm{\gamma} \in \R^{d}$ and denoting by $m$ the number of entries in the data set $\D_m$,

\begin{align}
    \beta_{j} =\sqrt{2\left\|\rho_{j}\right\|_{k}^{2}+300 \gamma_{j} \ln ^{3}\left(\frac{m+1}{1-\delta^{1/(d|\mathcal{N}|)}}\right)} 
\end{align}

The variable $\gamma_{j} \in \R$ is the maximum information gain
\begin{align}
    \gamma_{j} &=\max _{\bm{q}^{\{1\}}, \ldots, \bm{q}^{\{m+1\}} \in \Omega} \frac{1}{2} \log \left|I+\sigma_{j}^{-2} K\left(\x, \x^{\prime}\right)\right| \\
    \x, \x^{\prime} & \in\left\{\bm{q}^{\{1\}}, \ldots, \bm{q}^{\{m+1\}}\right\}.
\end{align}
\end{lemma}
\textit{Proof:} The results is a direct consequence of Lemma $1$ given in~\cite{beckers2019stable}.

\begin{remark}
An efficient algorithm can be used to find $\bm{\beta}$ based on the maximum information gain~\cite{srinivas2012information}.
\end{remark}

\section{Learning formation control based on GPR}\label{sec4}
Before proposing the learning-based formation control law, we revisit some results on the dynamical equations we will consider. Denote by $V$ the potential function $V=V_0+V_1:\R^{2d|\mathcal{N}|}\to\R$ given by  \begin{align}
	V(\bm{p}, \bm{v}) = \frac{1}{2}\sum_{i=1}^{|\mathcal{N}|} ||\bm{v}_i||^2 + \frac{1}{4}\sum_{k=1}^{|\mathcal{E}|}(||\bm{z}_{k}||^2-d_{k}^2)^2.
	\label{eq: phi}
\end{align}
In the absence of unknown disturbances, $V$ allows to write the closed-loop system  \cref{formation} as the system
\begin{align}
	\begin{cases}
	\dot{\bm{p}} = \nabla_\bm{v} V \\
	\dot{\bm{v}} = -\mathcal{K}\nabla_\bm{v} V -\nabla_\bm{p} V,
	\end{cases}
	\label{eq: H}
\end{align}
see~\cite{ahn14} for more details. Local asymptotic convergence and local exponential convergence for the system \cref{eq: H} to the formation \cref{S}  given by $\mathcal{S}$ with velocities of agents also driven to zero has been explored in \cite{ahn14}, \cite{sun2015rigid} and \cite{de2017taming}, by analyzing an equivalent decoupled gradient systems, all of them based on a result of \cite{dorfler2011critical}, that we will also use for the design of the online, decentralized learning-based control law and described as follow.\\
Consider the following one-parameter family of systems with double integrator formation stabilization dynamics $\mathcal{H}_\lambda$ given by
\begin{align}
\begin{bmatrix}\dot{\bm{p}} \\ \dot{\bm{v}}\end{bmatrix} = 
\begin{bmatrix}-\lambda I_{d|\mathcal{N}|} & (1-\lambda)I_{d|\mathcal{N}|} \\
	(\lambda-1)I_{d|\mathcal{N}|} & -\mathcal{K}I_{d|\mathcal{N}|}
\end{bmatrix}\begin{bmatrix}\nabla_\bm{p} V \\ \nabla_\bm{v} V \end{bmatrix},
	\label{eq: Hl}
\end{align}
where $\lambda \in [0, 1]$. Equation \cref{eq: Hl} continuously interpolates all convex combinations between the dissipative system (\ref{eq: H}) for $\lambda=0$ and a gradient system for $\lambda=1$. The family $\mathcal{H}_\lambda$ has two important properties summarized in the following Lemma from \cite{dorfler2011critical}.
\begin{lemma}\cite{dorfler2011critical}
	\label{lem: H}
\begin{itemize}
\item For all $\lambda \in [0, 1]$, the equilibrium set of $\mathcal{H}_\lambda$ is given by the set of the critical points of the potential function $V$, and is independent of $\lambda$.
	\item For any equilibrium of $\mathcal{H}_\lambda$ for all $\lambda \in [0, 1]$, the numbers of the stable, neutral, and unstable eigenvalues of the Jacobian of $\mathcal{H}_\lambda$ are the same and independent of~$\lambda$.
\end{itemize}
\end{lemma}
Denote by $E_{\bm{e},\bm{v}}:=(\bm{e},\bm{v})$ the stacked vector of relative positions errors and velocities for the formation stabilization, that is, the error in relative positions \cref{e_k} and velocities to achieve the desired formation described by \cref{S}. The next theorem introduces the learning-based control law with guaranteed boundedness of the error for the formation stabilization.
\begin{theorem}\label{theo:main}
Consider the system of agents~\cref{double2} with unknown dynamics and GP models with data sets~\cref{for:dataset} satisfying~\cref{ass:1,ass:2}. Assume that the desired formation shape $\mathcal{S}$ given by \cref{S} is infinitesimally and minimally rigid. Then, the control law \begin{align}\label{control-law}
    \bm{u}(t)=-\mathcal{K}\bm{v}-R^\top(\bm{z})\bm{e}(\bm{z})-\hat{\f}(\bm{p},\bm{v})-\bm{\mu}(\bm{\rho}|\bm{q},\D_m)
\end{align} guarantees that the error in the convergence to the desired shape $\mathcal{S}$ with zero velocity for all the agents, is uniformly ultimately bounded in probability by
\begin{align}
    \Prob\{||E_{\bm{e},\bm{v}}(t)||\leq\sqrt{2}\max_{q\in\Omega}\bar{\Delta}_{m(T)}(\bm{q}),\forall t\geq T_\delta\}\geq \delta
\end{align} on $\Omega$ with $T_\delta\in\R_{\geq 0}$. 
\end{theorem}
Note that the individual control law $\bm{u}_i(t)$ of each agent depends on the distance to its neighbors and the data set based on its own dynamics only.

\textit{Proof}: Consider the squared distance error for the edge $\mathcal{E}_k$, that is, $\bm{e}_k = ||\bm{z}_k||^2 - d^2_k$ and the stacked vector of squared distance errors denoted by $\bm{e} = \left[\bm{e}_{1}, \ldots, \bm{e}_{|\mathcal{E}|}\right]^{\top}$. Note that the time derivative of $\bm{e}_k$ is given by $\dot{\bm{e}}_k = 2\bm{z}_k^T\dot{\bm{z}}_k$.
Denoting by $E_{\bm{e},\bm{v}}^{\lambda}$ the stacked vector of errors $E_{\bm{e},\bm{v}}$ from~\cref{eq: Hl} for any $\lambda\in[0,1]$, which includes the closed-loop system~\cref{formation} for $\lambda=1$, we know that as a consequence of Lemma \ref{lem: H}, $E_{\bm{e},\bm{v}}^{\lambda}$ and $E_{\bm{e},\bm{v}}$ share the same stability properties. By using Lemma \ref{lem: H}, we will study the system (\ref{eq: Hl}) for $\lambda = 0.5$, without loss of generality, that is, 
\begin{align}
	\dot{\bm{p}} &= -\frac{1}{2}\overline BD(\bm{z})\bm{e} + \frac{1}{2}\bm{v} \label{eq:p} \\
	\dot{\bm{z}} &= -\frac{1}{2}\overline B^T\overline BD(\bm{z})\bm{e} + \frac{1}{2}\overline B^T \bm{v} \label{eq: zl} \\
	\dot{\bm{e}} &= -D(\bm{z})^T\overline B^T\overline BD(\bm{z})\bm{e} + D(\bm{z})^T\overline B^T \bm{v} \label{eq: el} \\
	\dot{\bm{v}} &= -\frac{1}{2}\overline BD(\bm{z})\bm{e} - \mathcal{K}\bm{v} \label{eq: vl}.
\end{align}
Consider the Lyapunov candidate function for the system \cref{double2} with control law $\bm{u}(t)$ given by \cref{control-law},
\begin{align}\label{V}
V(\bm{e},\bm{v})=\frac{1}{2}||\bm{e}||^2+||\bm{v}||^2.
\end{align} 
Note that $V$ is positive definite and radially unbounded. Next, we derive an upper bound for the time derivative of $V$. The time derivative of $V$ along the closed-loop system is given by
\begin{align}
    \dot{V}&=\bm{e}^\top\dot{\bm{e}}+2\bm{v}^\top\dot{\bm{v}}\\
    &=-\begin{bmatrix}\bm{e} & \bm{v} \end{bmatrix}\begin{bmatrix} D(\bm{z})^\top\overline B^T\overline BD(\bm{z}) & D(\bm{z})^\top\overline B^\top\\ \overline BD(\bm{z}) &  \mathcal{K}\end{bmatrix}\begin{bmatrix}\bm{e} \\ \bm{v} \end{bmatrix}+\bm{v}^\top(\bm{\rho}(\bm{q})-\bm{\mu}(\bm{\rho}|\bm{q},\D_m))\notag\\
    &=-\bm{e}^\top R(\bm{z})R(\bm{z})^\top\bm{e}-\bm{v}^\top\mathcal{K}\bm{v}+\bm{v}^\top(\bm{\rho}(\bm{q})-\bm{\mu}(\bm{\rho}|\bm{q},\D_m)).\notag
\end{align}

Denote by $\lambda_{min}$ and $\kappa_{min}$ the minimum eigenvalues of $R(\bm{z})R^\top(\bm{z})$ and $\mathcal{K}$, respectively. Since $\mathcal{S}$ is infinitesimally and minimally rigid then the rigidity matrix is full rank except the non-generic cases,  e.g., collinear or coplanar alignments of the agents in $\R^2$ or $\R^3$. Therefore $\lambda_{min}>0$. Note also that, as it was defined in \cref{formation}, $\kappa_{min}>0$. Therefore, by employing Lemma \ref{model-error} it follows that
\begin{align}\label{for:Lyapevo}
  \Prob\{\dot{V}\leq-\lambda_{\min}||\bm{e}||^2-\kappa_{\min}||\bm{v}||^2+||\bm{v}||\bar{\Delta}_m(\bm{q})\}\geq\delta,  
\end{align}
where $\bar{\Delta}_m(\bm{q}):\Omega\to\R_{\geq 0}$ is a bounded function such that $\Vert\bm{\beta}^{\top} \Sigma^{\frac{1}{2}}(\bm{\rho}\mid \bm{q},\D_m)\Vert\leq\bar{\Delta}_m(\bm{q})$, which exists because the kernel function is continuous and therefore it is bounded on a compact set $\Omega\subset\R^{2d|\mathcal{N}|}$, and then the variance $\Sigma(\bm{\rho}\mid \bm{q},\D_m)$ is bounded, see \cite{beckers2016equilibrium}.\\
Then, the value of $\dot{V}$ is negative with probability $\delta$ for all $E_{\bm{e},\bm{v}}$ with $\displaystyle{||E_{\bm{e},\bm{v}}||>\max_{\bm{q}\in\Omega}\sqrt{2}\bar{\Delta}_m(\bm{q})}$, where the maximum exists since $\bar{\Delta}_m(\bm{q})$ is bounded in $\Omega$. Finally, using~\cref{ass:1}, we define $T\in\R_{\geq 0}$ such that $\D_{m(T)}=\D_{m(t)}$ for all $t\geq T$. Then, $V$ is uniformly ultimately bounded in probability by $\displaystyle{\Prob\{||E_{v,e}||\leq b,
\,\forall t\geq T_\delta\in\R_{\geq 0}\}\geq\delta}$ with bound $\displaystyle{b=\max_{\bm{q}\in\Omega}\sqrt{2}\bar{\Delta}_{m(T)}(\bm{q})}$.\hfill$\square$

\begin{remark}
Depending on prior knowledge about the unknown function $\bm{\rho}$ the prediction error can vanish which leads to local asymptotic stability \cite{ahn14} if the framework describing the formation is rigid (resp, local exponential stability \cite{de2017taming} if the framework is infinitesimally and minimally rigid). In order to achieve this, we must assume that a perfect model was available, that is $\bar{\Delta}=0$. Then, from the computation of $\dot{V}$ one deduce the local asymptotic (resp., local exponential) stability for the formation stabilization problem. Simply speaking, to achieve this, the GP must be able to reproduce the unknown dynamics with a certain probability without any prediction error. This might be possible for certain types of unknown dynamics that are element of a RKHS spanned by a kernel with finite dimensional feature space, for instance, by the linear or the polynomial kernel, see~\cite{rasmussen2006gaussian}.
\end{remark}
\begin{remark}
In contrast to~\cite{yang2021distributed}, we consider no communication between the agents and our proposed control law in~\Cref{theo:main} allows i) online learning and ii) heterogeneous dynamics for the agents.
\end{remark}

\section{Numerical example}\label{sec5}
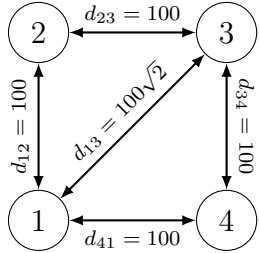
\begin{figure}[tbh]
\begin{center}
\tikzsetnextfilename{figure_graph}
	\begin{tikzpicture}[node distance=2.5cm,auto,>=latex,font={\sffamily}]
	\tikzstyle{block}=[draw,circle, minimum height=0.8cm,minimum width=0.8cm,inner sep=1pt]
    \node [block] (a1) {\large $1$};
    \node [block, above of=a1] (a2) {\large $2$};
    \node [block, right of=a2] (a3) {\large $3$};
    \node [block, right of=a1] (a4) {\large $4$};

    \draw[dedge] (a1) to node[rotate=90,anchor=south]{\footnotesize $d_{12}=100$} (a2);
    \draw[dedge] (a2) to node[]{\footnotesize $d_{23}=100$} (a3);
    \draw[dedge] (a3) to node[rotate=-90,anchor=south]{\footnotesize $d_{34}=100$} (a4);
    \draw[dedge] (a4) to node[]{\footnotesize $d_{41}=100$} (a1);
    \draw[dedge] (a1) to node[rotate=45,anchor=south]{\footnotesize $d_{13}=100\sqrt{2}$} (a3);
    %
\end{tikzpicture}
	\caption{Neighbor’s relations and desired shape}\vspace{-0.4cm}
	\label{fig:graph}
\end{center}
\end{figure}

In this section, we present a numerical example\footnote{A video of the simulation is available here: \url{https://youtu.be/WG8O8utmthA}} to evaluate the proposed control law. We consider $n=4$ agents in a $d=2$ dimensional space such that the position of each agent $i\in\mathcal{N}$ is denoted by $\bm{p}_i=[x_i,y_i]^\top$. The neighbor’s relations and desired shape are depicted in~\cref{fig:graph}. The graph defines a framework which is infinitesimally and minimally rigid, see~\cref{sec2b}. 
The double integrator dynamics~\cref{double} of agent 1 and agent 3 are affected by an arbitrarily chosen unknown dynamics
\begin{align}\label{for:f_ukn}
    \f_1(\bm{p}_1,\bm{v}_1)&=[200\sin(0.05 p_{1,y}), 200\cos(0.05 p_{1,x})]^\top\notag\\
    \f_3(\bm{p}_3,\bm{v}_3)&=[50\exp(-0.1(p_{3,y}-600)^2), 100]^\top,
\end{align}
respectively. The gain matrix $\mathcal{K}$ of the proposed control law~\cref{control-law} is set to $\mathcal{K}=2I_{2d|\mathcal{N}|}$. The GP models to predict~$\f_1,\f_3$ are equipped with a squared exponential kernel, see~\cite{rasmussen2006gaussian}. No prior model knowledge is assumed, i.e., $\hat{\f}_1=0,\hat{\f}_3=0$. At starting time $t=0$, the data set $\D_m$ is empty. A training point is added to $\D_m$ every $\SI{0.2}{\second}$ and the GP models are updated every $\SI{0.4}{\second}$ until a simulation time of $\SI{2.6}{\second}$. During each update of the GP model, the hyperparameters are optimized by means of the likelihood function, see~\cite{steinwart2008support}. We arbitrarily choose the following initial position $\bm{p}(0)=[450,450,510,610,590,590,650,550]^\top$.~\Cref{fig:pos1,fig:pos2} visualize the trajectories of the agents for the standard control law~\cref{formation} without GP model and the proposed control law~\cref{control-law} with GP model, respectively.
\begin{figure}[bth]
\begin{center}
\tikzsetnextfilename{figure_pos1}
	\begin{tikzpicture}
\begin{axis}[
name=plot1,
xlabel={x-Position},
  ylabel={y-Position},
  legend pos=north west,
  grid style={dashed,gray},
  grid = both,
       width=0.6*\columnwidth,
  height=5.3cm,
  axis equal,
  xtick={200,300,...,900},
  ytick={400,500,...,700},
  legend style={font=\footnotesize},
  legend pos= north east,
  legend cell align={left}]
\addplot[color=red,line width=1pt,no marks] table [x index=1,y index=2,col sep=comma]{data/data_pos_learning0.csv};
\addplot[color=red,line width=1pt,mark=triangle*,select coords between index={0}{0}] table [x index=1,y index=2,col sep=comma]{data/data_pos_learning0.csv};
\addplot[color=red,line width=1pt,mark=square*,select coords between index={299}{299}] table [x index=1,y index=2,col sep=comma]{data/data_pos_learning0.csv};
\addplot[color=blue,line width=1pt,no marks] table [x index=3,y index=4,col sep=comma]{data/data_pos_learning0.csv};
\addplot[color=blue,line width=1pt,mark=triangle*,select coords between index={0}{0}] table [x index=3,y index=4,col sep=comma]{data/data_pos_learning0.csv};
\addplot[color=blue,line width=1pt,mark=square*,select coords between index={299}{299}] table [x index=3,y index=4,col sep=comma]{data/data_pos_learning0.csv};
\addplot[color=green!80!black, line width=1pt,no marks] table [x index=5,y index=6,col sep=comma]{data/data_pos_learning0.csv};
\addplot[color=green!80!black,line width=1pt,mark=triangle*,select coords between index={0}{0}] table [x index=5,y index=6,col sep=comma]{data/data_pos_learning0.csv};
\addplot[color=green!80!black,line width=1pt,mark=square*,select coords between index={299}{299}] table [x index=5,y index=6,col sep=comma]{data/data_pos_learning0.csv};
\addplot[color=black, line width=1pt,no marks] table [x index=7,y index=8,col sep=comma]{data/data_pos_learning0.csv};
\addplot[color=black,line width=1pt,mark=triangle*,select coords between index={0}{0}] table [x index=7,y index=8,col sep=comma]{data/data_pos_learning0.csv};
\addplot[color=black,line width=1pt,mark=square*,select coords between index={299}{299}] table [x index=7,y index=8,col sep=comma]{data/data_pos_learning0.csv};
\legend{Agent 1,,,Agent 2,,,Agent 3,,,Agent 4}
\node[anchor=west] at (axis cs:200,760){No learning};
\end{axis}
\end{tikzpicture} 
	\vspace{-0.2cm}\caption{Trajectory with initial position (triangle) and final position (square) of the agents for the standard control law without learning.}\vspace{-0.4cm}
	\label{fig:pos1}
\end{center}
\end{figure}
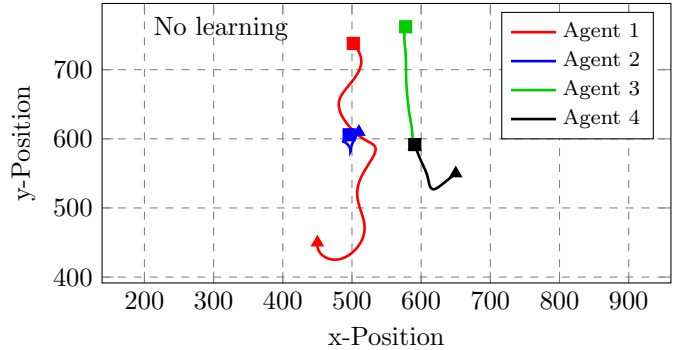
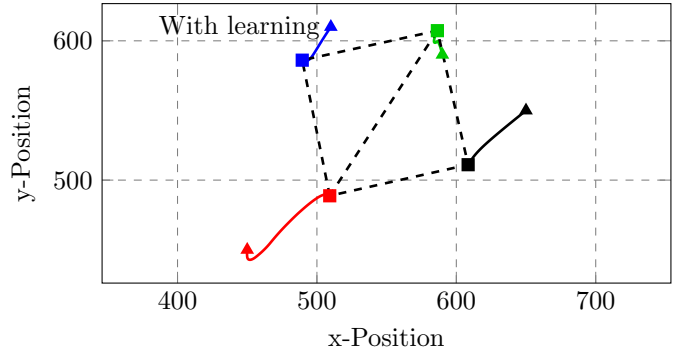
\begin{figure}[hbt]
\begin{center}
\tikzsetnextfilename{figure_pos2}
	\begin{tikzpicture}
\begin{axis}[
  xlabel={x-Position},
  ylabel={y-Position},
  legend pos=north west,
  grid style={dashed,gray},
  grid = both,
       width=0.6*\columnwidth,
  height=5.3cm,
  axis equal,
  xtick={400,500,...,700},
  ytick={500,600},
  legend style={font=\footnotesize},
  legend cell align={left}]
\addplot[color=red,line width=1pt,no marks] table [x index=1,y index=2,col sep=comma]{data/data_pos_learning1.csv};
\addplot[color=red,line width=1pt,mark=triangle*,select coords between index={0}{0}] table [x index=1,y index=2,col sep=comma]{data/data_pos_learning1.csv};
\addplot[color=red,line width=1pt,mark=square*,select coords between index={299}{299}] table [x index=1,y index=2,col sep=comma]{data/data_pos_learning1.csv};
\addplot[color=blue,line width=1pt,no marks] table [x index=3,y index=4,col sep=comma]{data/data_pos_learning1.csv};
\addplot[color=blue,line width=1pt,mark=triangle*,select coords between index={0}{0}] table [x index=3,y index=4,col sep=comma]{data/data_pos_learning1.csv};
\addplot[color=blue,line width=1pt,mark=square*,select coords between index={299}{299}] table [x index=3,y index=4,col sep=comma]{data/data_pos_learning1.csv};
\addplot[color=green!80!black, line width=1pt,no marks] table [x index=5,y index=6,col sep=comma]{data/data_pos_learning1.csv};
\addplot[color=green!80!black,line width=1pt,mark=triangle*,select coords between index={0}{0}] table [x index=5,y index=6,col sep=comma]{data/data_pos_learning1.csv};
\addplot[color=green!80!black,line width=1pt,mark=square*,select coords between index={299}{299}] table [x index=5,y index=6,col sep=comma]{data/data_pos_learning1.csv};
\addplot[color=black, line width=1pt,no marks] table [x index=7,y index=8,col sep=comma]{data/data_pos_learning1.csv};
\addplot[color=black,line width=1pt,mark=triangle*,select coords between index={0}{0}] table [x index=7,y index=8,col sep=comma]{data/data_pos_learning1.csv};
\addplot[color=black,line width=1pt,mark=square*,select coords between index={299}{299}] table [x index=7,y index=8,col sep=comma]{data/data_pos_learning1.csv};
\addplot[color=black, dashed, line width=1pt,no marks] coordinates{(5.0919e+02,4.8875e+02)
(4.8947e+02,5.8611e+02)
(5.8642e+02,6.0729e+02)
(6.0848e+02,5.1106e+02)
(5.0919e+02,4.8875e+02)
(5.8642e+02,6.0729e+02)};
\node[anchor=west] at (axis cs:380,610){With learning};
\end{axis}
\end{tikzpicture} 
	\vspace{-0.2cm}\caption{Trajectory with initial position (triangle) and final position (square) of the agents for the proposed control law with GP model where the agents converge to the desired formation.}\vspace{-0.4cm}
	\label{fig:pos2}
\end{center}
\end{figure}
The initial position of each agent is denoted by a triangle whereas the position after the simulation time of $\SI{6}{\second}$ is denoted by a square. The standard control approach fails to reach the desired formation as shown in~\cref{fig:graph}. In contrast, the GP models allow the agents to converge to a tight set around the desired formation.\\
The evolution of the Lyapunov function in~\cref{fig:V} highlights the superior of the proposed control law as it allows the Lyapunov function to converge to a tight set around zero. Note that the evolution of the Lyapunov function is not always decreasing but bounded in a neighborhood around zero, see~\cref{for:Lyapevo}. The size of the set shrinks with improved accuracy of the GP model. The online learning process for agent 1 and agent 3 is depicted in~\cref{fig:fukn_gpr}. The solid line represents the unknown dynamics~\cref{for:f_ukn} over time whereas the dashed line is given by the GP prediction. The jumps of the prediction occur due to the model update every $\SI{0.4}{\second}$. After $\SI{2}{\second}$, the GP model can accurately mimic the unknown dynamics.
\begin{figure}[bht]
\begin{center}
\tikzsetnextfilename{figure_V}
	\begin{tikzpicture}
\begin{axis}[
  ylabel={Lyapunov V},
  xlabel={Time [s]},
  legend pos=north west,
  grid style={dashed,gray},
  grid = both,
       width=0.6*\columnwidth,
  height=4cm,
  xmin=0,
  xmax=6,
  legend style={font=\footnotesize},
  legend cell align={left},
  legend pos = north east]
\addplot[color=black,dashed,line width=1pt,no marks] table [x index=0,y expr=\thisrowno{1}/5100,col sep=comma]{data/data_V_learning0.csv};
\addplot[color=black,line width=1pt,no marks] table [x index=0,y expr=\thisrowno{1}/5100,col sep=comma]{data/data_V_learning1.csv};
\legend{No learning, With learning};
\end{axis}
\end{tikzpicture} 
	\vspace{-0.2cm}\caption{Normalized Lyapunov function of the closed-loop with the standard control law (dashed) and the proposed, learning-based control law (solid) which converges to a tight set around zero.}\vspace{-0.7cm}
	\label{fig:V}
\end{center}
\end{figure}
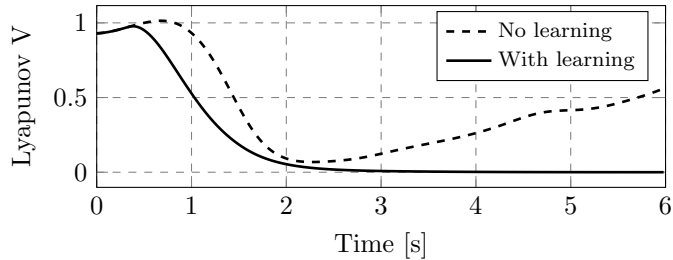
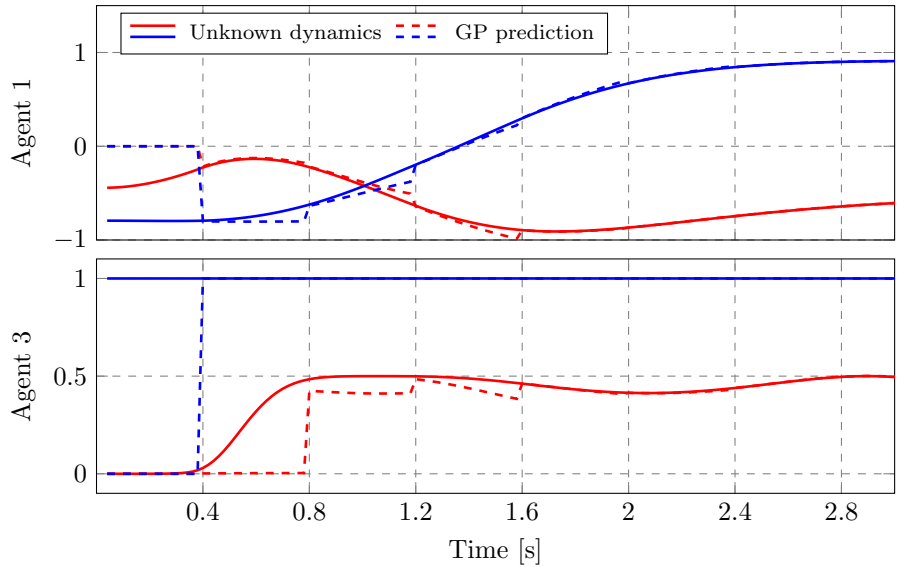
\begin{figure}[h]
\begin{center}
\tikzsetnextfilename{figure_fukn_gpr}
	\begin{tikzpicture}
\begin{axis}[
name=plot1,
  ylabel={Agent 1},
  legend pos=north west,
  grid style={dashed,gray},
  grid = both,
       width=0.8*\columnwidth,
  height=4.7cm,
  ymin=-1,
  ymax=1.5,
  xmin=0,
  xmax=3,
  xtick={0.4,0.8,...,2.8},
  xticklabels={},
  legend style={font=\footnotesize,row sep=-3pt,column sep=0.1cm},
    legend columns=2,
  transpose legend,
  legend cell align={left}]
\addplot[color=red,line width=1pt,no marks] table [x expr=\thisrowno{0}+0.04,y expr=\thisrowno{1}/220,col sep=comma]{data/data_fukn_gpr_learning1.csv};
\addplot[color=blue,line width=1pt,no marks] table [x expr=\thisrowno{0}+0.04,y expr=\thisrowno{2}/220,col sep=comma]{data/data_fukn_gpr_learning1.csv};
\addplot[color=red,dashed, line width=1pt,no marks] table [x expr=\thisrowno{0}+0.04,y expr=\thisrowno{3}/220,col sep=comma]{data/data_fukn_gpr_learning1.csv};
\addplot[color=blue,dashed, line width=1pt,no marks] table [x expr=\thisrowno{0}+0.04,y expr=\thisrowno{4}/220,col sep=comma]{data/data_fukn_gpr_learning1.csv};
\legend{\raisebox{-2mm}[0mm][0mm]{Unknown dynamics},\phantom{a}, \raisebox{-2mm}[0mm][0mm]{GP prediction},\phantom{a}};
\end{axis}
\begin{axis}[
name=plot2,
   at=(plot1.below south east), anchor=above north east,
  xlabel={Time [s]},
  ylabel={Agent 3},
  legend pos=north west,
  grid style={dashed,gray},
  grid = both,
       width=0.8*\columnwidth,
  height=4.7cm,
  xmin=0,
  xmax=3,
  xtick={0.4,0.8,...,2.8},
  legend style={font=\footnotesize},
  legend cell align={left}]
\addplot[color=red,line width=1pt,no marks] table [x expr=\thisrowno{0}+0.04,y expr=\thisrowno{9}/100,col sep=comma]{data/data_fukn_gpr_learning1.csv};
\addplot[color=blue,line width=1pt,no marks] table [x expr=\thisrowno{0}+0.04,y expr=\thisrowno{10}/100,col sep=comma]{data/data_fukn_gpr_learning1.csv};
\addplot[color=red,dashed, line width=1pt,no marks] table [x expr=\thisrowno{0}+0.04,y expr=\thisrowno{11}/100,col sep=comma]{data/data_fukn_gpr_learning1.csv};
\addplot[color=blue,dashed, line width=1pt,no marks] table [x expr=\thisrowno{0}+0.04,y expr=\thisrowno{12}/100,col sep=comma]{data/data_fukn_gpr_learning1.csv};
\end{axis}
\end{tikzpicture} 
	\vspace{-0cm}\caption{The x-component (red, solid) and y-component (blue, solid) of the unknown dynamics~\cref{for:f_ukn} as ground truth and the prediction of the GP model (dashed) for agent 1 (top) and agent 3 (bottom). The GP model is updated every $\SI{0.4}{\second}$ with new data.}\vspace{-0.4cm}
	\label{fig:fukn_gpr}
\end{center}
\end{figure}
\section*{Conclusion}
We propose a safe online learning-based formation control law for double-integrator agents with partially unknown dynamics. In this scenario, the agents can measure inter-agent distances to neighbors only and collect their own training data online. By using online updated Gaussian process models and the proposed decentralized control law, we prove that the error to desired formation shapes is uniformly ultimately bounded in probability. In addition, the size of the bound is explicitly given and shrinks for improved GP model accuracy. Finally, a numerical example with 4 agents visualizes the effectiveness of the control law.

\section*{Acknowledgements}
This work was supported by a $2020$ Leonardo Grant for Researchers and Cultural Creators, BBVA Foundation and by the European Research Council (ERC) Consolidator Grant ``Safe data-driven control for human-centric systems (COMAN)” under grant agreement number 864686. L. Colombo have been partially founded by MINECO grant MTM2016-76072-P and a fellowship from ``la Caixa' Foundation under fellowship code LCF/BQ/PI19/11690016.

\bibliographystyle{IEEEtran}
\bibliography{root}

\end{document}